\documentclass[11pt]{article}

\input epsf.sty

\def\lromn#1{\uppercase\expandafter{\romannumeral#1}}

\begin{document}
\begin{flushright}
\end{flushright}

\begin{center}
\begin{Large}
{\bf Neutrino mass determination using
circulating heavy ions
}
\end{Large}

\vspace{2cm}
M.~Yoshimura

\vspace{0.5cm}
Center of Quantum Universe, Faculty of
Science, Okayama University \\
Tsushima-naka 3-1-1 Kita-ku Okayama
700-8530 Japan

\end{center}

\vspace{5cm}

\begin{center}
\begin{Large}
{\bf ABSTRACT}
\end{Large}
\end{center}

We study process of radiative neutrino pair emission
$|e\rangle \rightarrow |g \rangle + \gamma +\nu \bar{\nu}$
from coherently excited heavy ions 
(quantum mixture of two ionic states, the ground and an excited states) in circular motion. 
Determination of the neutrino mass is found to be possible with
simultaneous detection of the photon and  one of neutrinos in the pair
down to the level  of the smallest neutrino mass of order 5 meV
 in the three flavor scheme.

\vspace{4cm}
PACS numbers
\hspace{0.5cm} 
13.15.+g, 
14.60.Pq, 

Keywords
\hspace{0.5cm} 
neutrino mass,
heavy ion synchrotron,
quantum coherence
\newpage

\section
{\bf Introduction}

Conventional neutrino sources have been weak decay products of 
elementary particles (mainly pion and muon) and
$\beta-$nuclei.
It was recently proposed \cite{pair beam} that circulating excited
heavy ions   may become another unique  source producing large amounts of pairs of
neutrino and anti-neutrino of the same flavor $\nu_a \bar{\nu}_a, a = e, \mu, \tau$.
Quantum mixture of excited and ground ionic states,
 realized by  laser irradiation from counter-propagating directions
against ions,
inputs an ionic internal energy  into the orbital motion,
leading to a new kind of non-linear resonance.
This resonance gives for rate calculations a stationary point in the crucial time integral of
the phase factor, which  otherwise  
monotonically varies, hence giving much smaller rates  with much lower neutrino energies, 
as in the case of synchrotron radiation \cite{schwinger}.
When the  quantum number of excited level is
appropriately chosen, neutrino pairs may be emitted with large rates
 during the resonance decay time.
Produced neutrino energies may reach
the GeV region suitable for oscillation experiments.

In the previous work \cite{nu-oscillation in pair beam}  
physics related to the pair emission process $|e \rangle \rightarrow |g \rangle  +\nu \bar{\nu}$
was discussed.
In the present work, we examine a different process:  radiative emission of neutrino pair  (RENP)
$|e \rangle \rightarrow |g \rangle + \gamma +\nu \bar{\nu}$
from circulating excited heavy ions.
Relative parities between the state $|e\rangle $ and the state $|g\rangle$ in this case 
are different ($-$), while they are the same ($+$) in the case of non-radiative neutrino pair emission.
Thus, two processes cannot occur simultaneously.
In order to avoid strong electric dipole transition (a purely QED process) against RENP,
it is necessary to have a large angular momentum change like $|\Delta J| \geq 2$
between two ionic states,  $|e\rangle $ and  $|g\rangle$. 
A candidate ionic state for $|e \rangle$ is Ne-like heavy ion states in
the configuration of $(2p^5 3s)_{J=2}$, for instance Pb$^{72+}$ giving
the first metastable state of $J^P=2^-$ of energy $O(1)$keV.
In this work we shall not attempt a serious search for the best candidate ion, since
this requires both of detailed experimental R and D works related to acceleration
and dedicated atomic physics calculation.
We shall concentrate on experimental principles and
rate estimates assuming a hopefully reasonable set
of parameter values.

The rest of this paper is organized as follows.
In the next section we recapitulate the main features of
beam RENP from 
quantum coherent heavy ion beam, offering an intuitive
understanding in terms of the non-linear resonance.
Section 3 is the main part of the present work and
provides calculation of RENP spectrum rates, both when neutrino variables are
integrated out and when one of neutrino variables is left for its detection.
Section 4 addresses the question of the sensitivity to
the smallest neutrino mass in the three flavor scheme
and discusses how small one can determine the smallest
neutrino mass taking the case of normal
hierarchical (NH) mass pattern..

Throughout this work we use the natural unit of $\hbar = c = 1$.

\vspace{0.5cm}
\section {\bf Resonance condition of coherently excited ion in circular motion}

We first recapitulate the important feature of beam RENP.

A promising way to prepare a coherent  state of
excited and ground states is via 
laser irradiation, often more than two lasers for neutrino pair emission.
It may lead to a quantum mixture of two states,
\begin{eqnarray}
&&
| c \rangle = \cos \theta_c |g \rangle + \sin \theta_c e^{i\varphi_c} | e\rangle
\,,
\end{eqnarray}
where the angle and the phase, $\theta_c$ and $\varphi_c$, may vary in time, 
but we consider a shorter time
scale than this variation time.
Without any phase relaxation (a reasonable assumption after the
circulating beam energy reaches its maximum), this state maintains its quantum nature,
and various observable quantities are given in terms of the density matrix
$ |c \rangle \langle c|$ for the pure quantum state.
Its off-diagonal element $\rho_{eg} = \sin (2\theta_c) e^{-i\varphi_c}/2$
is called the coherence.
An experimental method (adiabatic Raman excitation) realizing a high coherence is described
and was successfully applied in \cite{psr observation} to gain
a high level of quantum coherence of several \%
in a  macroscopic system of hydrogen molecules in a gas phase.

Unlike synchrotron radiation \cite{schwinger},
the crucial time integral of phase factor $e^{i\Phi(t)}$
contains two contributions: the one coming from the
ion circular motion and the other from the internal transition.
These two terms may have phases of different signs in the time integral, and there is
a possibility of phase cancellation.
The cancellation is interpreted as a kind of resonance,
and the resonance condition is given by the stationarity condition, $\partial_t \Phi = 0$.
Emission of photon of energy $\omega$ and
neutrino pair of energies $E_i, i = 1,2$, in RENP gives the resonance condition,
\begin{eqnarray}
&&
(\omega + E_1 + E_2 )(1 - \cos \frac{vt}{\rho} ) = \frac{\epsilon_{eg}}{\gamma}
\,,
\label {resonance condition}
\end{eqnarray}
when  all three particles $\gamma, \nu, \bar{\nu}$ are assumed to be emitted 
at the forward 
direction.
Since the angular distribution is well collimated around the forward direction,
this is a reasonable assumption at least for an estimate.
The constant $\gamma$ here
is the boost factor determined by a circular velocity $v$: $\gamma = 1/\sqrt{1-v^2}$.
The radius of the circular motion  denoted here by $\rho$
is equal to the Zeeman splitting energy $eB/(M \gamma)$
under a magnetic field $B$ for circular motion.
The time $t$ here is the look-back time measured from a fixed present time $t=0$, and 
for $t \ll \rho$ the resonance formation time $t_r$ is calculated from (\ref{resonance condition}) as
\begin{eqnarray}
&&
t_r \sim \rho \sqrt{\frac{2 \epsilon_{eg} }{\gamma(\omega + E_1 + E_2) }}
\,.
\end{eqnarray}
The RENP photon spectrum is continuous and the
total energy of three particles is bounded from above:
 $\omega + E_1 + E_2 \leq 2 \epsilon_{eg} \gamma$.
This gives $t_r \geq \rho/\gamma$.
Numerically, taking the total energy of order its maximum $2 \epsilon_{eg}\gamma$, 
$t_r = \rho/\gamma \sim 3\times 10^{-10}$s for $\rho = 4$km
and $\gamma = 5\times 10^4$.
Dominant contribution to the phase integral comes from the resonance region, and
this integral is approximately $\int_0^{\infty} dt  \cos \left( (t - t_r)^2/(2 \Delta t_r^2)\right) $.
The width factor $\Delta t_r \sim \sqrt{\rho/\epsilon_{eg}}$ is of order $3 \times 10^{-12}$s
for $\epsilon_{eg} = 1 {\rm keV}, \rho = 4$km, namely
the resonance is very sharp, $ \Delta t_r \ll t_r$, in the time domain,
and is very broad in the energy domain.
The resonance shape in the energy domain is not of a simple Lorentzian type.

More precisely, the resonance behavior of the phase integral
is worked out, using the technique of \cite{pair beam},
by the formula,
\begin{eqnarray}
&&
\int_0 ^{\infty} dt \cos \Phi (t)
\,, \hspace{0.5cm}
\Phi(t) =\frac{
\omega+ 
E_1 + E_2}{2\rho \gamma} \sqrt{D}
( t - \frac{\rho}{\gamma} D )^2
\,, \hspace{0.5cm} 
D =  \frac{{\cal E}}{\omega+E_1 + E_2 } 
\,,
\label {angular constraint eq 0}
\\ &&
{\cal E} = \frac{\epsilon_{eg}}{\gamma} - \omega -E_1 - E_2 +
(1 - \frac{1}{\gamma^2})^{1/2}
(\omega \cos \psi \cos \theta + \sum_i p_i \cos \psi_i \cos \theta_i )
\nonumber \\ &&
+ \frac{1}{2}(1 - \frac{1}{\gamma^2})^{1/2}
\frac{ (\omega \cos \psi \sin \theta + \sum_i p_i \cos \psi_i \sin \theta_i )^2}
{ \omega \cos \psi \cos \theta + \sum_i p_i \cos \psi_i \cos \theta_i}
\\ &&
\hspace*{-0.5cm}
\sim 
\frac{ \epsilon_{eg} }{ \gamma}
-
\frac{ m_i^2/ E_1 + m_j^2/E_2}{\gamma^2}
-
\frac{\omega }{2\gamma^2}(\theta^2 + \psi^2) -
\sum_i 
\frac{E_i }{2\gamma^2}(\theta_i^2 + \psi_i^2)
+ \frac{1}{2} \frac{( \omega\theta  + \sum_i E_i \theta_i)^2 }
{\omega +E_1 + E_2}
- \frac{\omega +E_1 + E_2} {2\gamma^2 }
\,,
\label {angular constraint eq}
\end{eqnarray}
for a neutrino pair emission of masses, $m_i, m_j$.
Emission angles ($\psi, \theta $ for the photon, $\psi_i, \theta_i\,, i=1,2$ for the neutrino pair)
are defined for the forward direction  along
the circulating beam to be at $\theta = \psi = 0$.
For simplicity we took the small  neutrino mass limit,
which should be sufficient for our consideration of much
larger neutrino energies.
In this Airy-type of integral
the development time $t_r$ and the resonance width $\Delta t_r$  are
\begin{eqnarray}
&&
{\rm resonance \; in \; time \; domain}:
\hspace{0.3cm}
t_r = \frac{\rho}{\gamma} D 
\,, \hspace{0.5cm}
\\ &&
{\rm width };
\hspace{0.3cm}
\Delta t_r = \rho \sqrt{ \frac{2 }{(\omega+ E_1 +E_2) t_r }} \gg t_r
\,,
\end{eqnarray}
giving
\begin{eqnarray}
&&
\int_0 ^{\infty} dt \cos \Phi (t) \sim \int_0 ^{\infty} dt \cos \frac{( t -t_r)^2 }{(\Delta t_r )^2 }
\sim \sqrt{\frac{ 2\pi}{3 }}\Delta t_r = 
\sqrt{\frac{ \pi}{3}} (\frac{\rho \gamma }{\omega+E_1+E_2 })^{1/2} D^{-1/4}
\,.
\end{eqnarray}
The allowed range of dimensionless function $D$ is $0 \sim 1$.
The restriction $D \geq 0$ along with positivity of variables gives
constraints on angular variables and energies of all three particles.

It is necessary to irradiate lasers from the counter-propagating
directions against the circulating beam at lease once in
each revolution of circular motion,  in order to efficiently 
re-excite lost available heavy ions.
Moreover, irradiation from the counter-propagating directions effectively boosts 
laser frequencies in optical and infrared regions into the keV range of X-rays.

\vspace{0.5cm}
\section {\bf RENP spectrum rates from circulating excited ions}

The RENP idea 
has originally been developed for  small scale laboratory experiments aiming at systematic exploration 
of unknown neutrino properties \cite{renp overview}.
The macro-coherent amplification (without the wavelength limitation)
of otherwise tiny weak rates
in atomic processes is the key concept  for success of this project.
This way one may effectively enhance weak rates by
a factor $\propto n^2$ where $n$ is the number density of atoms in the
available level.
We recently observed the macro-coherent amplification in
weak QED processes \cite{psr observation}, but there are still many steps
towards the final goal.

In the present work
we shall instead employ quantum coherence at a single ion level:
the macro-coherence amplification is not necessary.
But if it exists, it certainly helps much in giving much larger rates than
given here.
Acceleration of coherently excited ion in the present scheme is a new  innovation  and
would require much R and D works, but
it has a potential of producing coherent gamma ray beam
if a macro-coherence is realized
\cite{pair beam}, which may have rewarding applications.

We now work out RENP spectrum rates  from circulating excited ions.
The process for a single ion is
$|e \rangle \rightarrow |g \rangle + \gamma + \nu_i \bar{\nu}_j\,, i= 1,2,3$,
where $ \nu_i$ and $\bar{\nu}_j$ are mass eigen-states (in the Majorana case
$ \nu_i = \bar{\nu}_i$).
The important  angular variable dependence of emitted neutrinos 
is in $D$, which is to a good approximation
quadratic in angular variables as shown in eq.(\ref{angular constraint eq}).
One can integrate over these angular variables by rescaling their
magnitudes by $1/\gamma$, resulting  in the following equation (\ref{integrals}).
The photon energy spectrum when two neutrino variables are integrated out
is then calculated in a convenient integral form of two neutrino energies
 scaled by the maximum total energy $\omega_m$, $x_i =E_i/\omega_m,  i = 1,2$,
\begin{eqnarray}
&&
\omega_m \frac{d\Gamma_{ij}}{d\omega} = R F_{ij}(\frac{\omega}{\omega_m})
\,, \hspace{0.5cm}
\omega_m = 2 \epsilon_{eg} \gamma
\,, 
\\ &&
R = 
\frac{\sqrt{\pi}}{ 2 \sqrt{3}  (2\pi)^8  } v_5
G_F^2  d_{pe}^2 \gamma^6 N |\rho_{eg}(0)|^2
\sqrt{\rho} \epsilon_{eg}^{19/2}
\frac{1}{\epsilon_{pe}^2} 
\nonumber
\\ &&
\sim 
1.46 \times 10^{12} {\rm Hz}
\frac{ N|\rho_{eg}(0)|^2}{10^8} 
\frac{ \gamma_{pe}}{ 100 {\rm MHz}}
\sqrt{\frac{\rho}{4 {\rm km}}}
(\frac{\gamma}{10^{4}})^6
(\frac{\epsilon_{eg}}{1 {\rm keV}})^{15/2}
\,, 
\label {overall rate}
\\ &&
v_5 = \int dV_5 ( 1 - r^2)^{-1/4} \sim 9.1 \times 10^{-6}
\,, \hspace{0.5cm}
F_{ij}(y) = \int_{0}^1 dx_1 \int_{0}^1dx_2 H_{ij}(y, x_1, x_2)
\,,
\label {integrals}
\\ &&
\hspace{-0.5cm}
H_{ij}(y, x_1, x_2) = y^{5/2} (1 + \frac{2\epsilon_{eg}}{\epsilon_{pe}} y)^{-2}
x_1 x_2(x_1 + x_2 + y)^{1/4}
G_{ij}(x_1,x_2, y)^{9/4}
\Theta(G_{ij}(x_1,x_2, y)\, )
\,,
\\ &&
G_{ij}(x_1,x_2, y) = 1 - x_1 -x_2 - y - \frac{1}{4 \epsilon_{eg}^2}
(\frac{m_i^2}{x_1} + \frac{m_j^2}{x_2})
\,,
\end{eqnarray}
where $|p \rangle$ is an ionic state having a higher excitation energy than $|e \rangle$.
In the example of Ne-like ion of $|e\rangle = (2p^5 3s)_{J=2}$ one of these states may be
$|p\rangle = (2p^5 3p)_{J=1}$.
We assumed $\epsilon_{pe} =\epsilon_{eg}/10$ (close to Ne-like ion level spacing for this type of $|p \rangle$)
for simplicity.
The dipole moment $d_{pe}$ is related to the E1 transition rate $\gamma_{pe}$
which we assume of order 100 MHz.
The step function $\Theta(G)$ is defined by the property:
$=1$ for $G >0 $ and  $=0$ for $G <0 $.
The neutrino energy range in the integral here
is determined by $G \geq 0$ and $x_i \geq m_i/(2\epsilon_{eg}\gamma)$.

In this formula we incorporated finite neutrino 
mass effects only kinematically, hence there is no rate difference between the Majorana and
Dirac neutrinos.
The matrix element factor distinguishes Majorana and Dirac  cases
and shall be treated below.
The boost factor ($\gamma$) dependence is different  from the case of neutrino-pair beam, 
and  is larger by $\gamma^2$
than the previous  case, as expected from a $\gamma-$ scaling law \cite{pair beam}.

\begin{figure*}[htbp]
 \begin{center}
 \epsfxsize=0.5\textwidth
 \centerline{\epsfbox{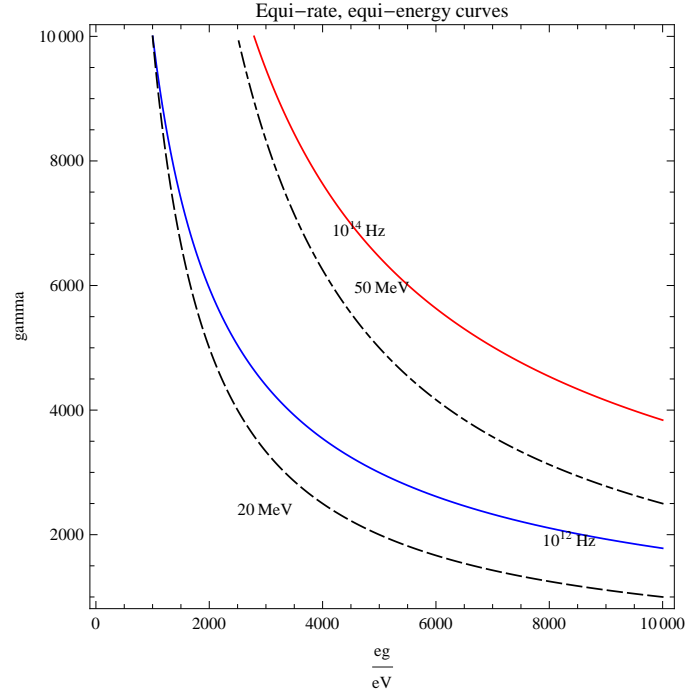}} \hspace*{\fill}
   \caption{RENP rate contours given by the overall factor $R$
of eq.(\ref{overall rate}) and equi-$\omega_m$ contours in $(\epsilon_{eg}, \gamma)$ plane.
$\epsilon_{pe}/\epsilon_{eg} = 1/2, \gamma_{pe} = 100 $MHz assumed.
}
   \label {renp kev rate contour}
 \end{center} 
\end{figure*}

\begin{figure*}[htbp]
 \begin{center}
 \epsfxsize=0.6\textwidth
 \centerline{\epsfbox{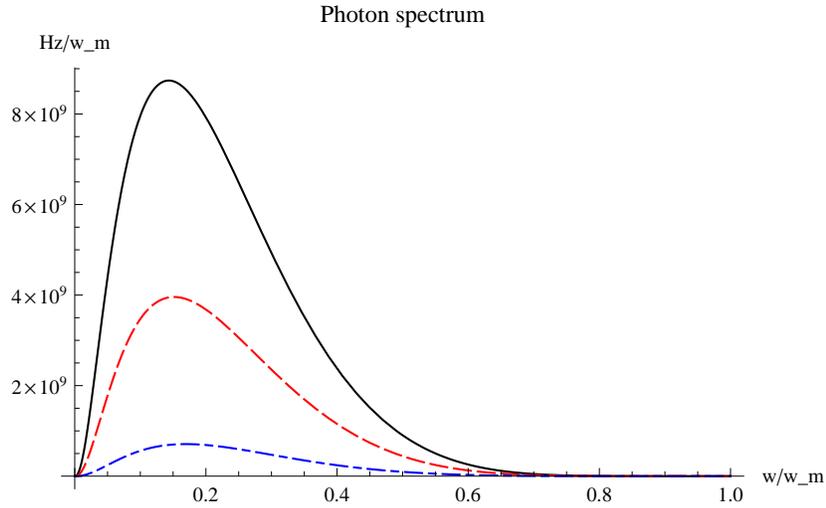}} \hspace*{\fill}
   \caption{Photon energy spectral rates for a massless neutrino:
$\epsilon_{eg}/\epsilon_{pe} =$ 10 in solid black, 8 in dashed red, and
5 in dash-dotted blue.
Other assumed common parameters are
$N|\rho_{eg}(0)|^2 = 10^8,
\gamma_{pe}= 100 {\rm MHz},
\rho =4 {\rm km}$, 
$\epsilon_{eg} =1 {\rm keV}, \gamma = 5 \times 10^4$.
}
   \label {massless rates}
 \end{center} 
\end{figure*}

Without any experimental R and D works, it is difficult to estimate the number of
excited heavy ions with a high coherence denoted by the parameter $N|\rho_{eg}(0)|^2$.
We took  $10^8$ for this value under this circumstance.
This is the most important uncertainty in rate estimates of the present work.
The overall RENP rate factor $R$ of eq.(\ref{overall rate})
in the $(\epsilon_{eg}, \gamma)$ plane are plotted in Fig(\ref {renp kev rate contour}).
and Fig(\ref{massless rates}) we illustrate photon spectral rates
for a single massless neutrino pair emission, showing their sensitivity to the energy ratio
$\epsilon_{pe}/\epsilon_{eg}$.
We may judge from these figures that RENP from circulating coherent heavy ion
can be measured with high enough rates, if  $N|\rho_{eg}(0)|^2 $ is not too small.

A term 
\begin{eqnarray}
&&
- \frac{3}{8} \frac{m_i m_j} {E_i E_j} \Re(c_{ij}^2)
= - \frac{3}{8} \frac{m_i m_j} {\omega_m^2 x_1 x_2} \Re(c_{ij}^2)
\,, \hspace{0.5cm}
 c_{ij} = U_{ei}^* U_{ej} - \frac{1}{2} \delta_{ij}
\,.
\label {mixing factor}
\end{eqnarray}
is added in the Majorana case to squared matrix elements  common to Dirac and Majorana terms,
$|c_{ij}^2|$, in the pair emission of mass sigenstates $(i,j)$,
This term arises from anti-symmetrized wave functions of two neutrinos,
only present for identical Majorana fermions \cite{my-prd-07}.
Here $ U_{ei}, i =1,2,3$, is the $3\times 3$ unitary matrix elements describing the neutrino mixing, and
its elements have been determined
from oscillation experiments   \cite{mixing parameters} except one (in the Dirac case) 
or three (in the Majorana case) phase factors.
The important difference from RENP of SPAN project \cite{renp overview} is
that the conservation laws of energy and momentum 
do not hold, and contributions near the mass thresholds
$E_i \approx m_i$ can exist  for  any given photon energy $\omega$.
It is however found in the present work  that 
the neutrino mass suppression given by
$m_{i}/E_{i}$ is severe at high neutrino energies $E_i$
for the Majorana/Dirac distinction.

Non-conservation of energy and momentum actually means that
non-conserved amounts are compensated by
 ion de-excitation and 
its recoil, which are a small portion of their total amounts.
It is however important to recover these by re-pumping ions
with laser irradiation from counter-propagating direction
at a straight section of the synchrotron machine.
This way one justifies the method of rate calculation presented here.

The main background when only a photon is detected
is the three photon decay in which two photons escape detection.
This background is rejected to some extent by
subtracting escaped three photon events.
The subtraction  may be effective by different spectral shapes
in the two processes.
Still, there might be substantial background coming in
after the subtraction.
In this case one would have to rely on
parity violating quantities such as emergent
circular polarization that exist only in weak interaction
process  in order to unambiguously identity RENP.
One needs a detailed study of the background rejection  based on simulations
taking into account a detector design.

\vspace{0.5cm}
\section {\bf Neutrino mass determination and comparison with
SPAN}

An efficient determination of the smallest neutrino mass $m_0$
requires a more refined measurement.
It indeed becomes possible to determine the smallest neutrino mass,
if one of pair neutrinos is simultaneously detected along with a photon.
The double energy spectrum of $\gamma+\nu_i$ is given by
\begin{eqnarray}
&&
\omega_m^2 \frac{d^2\Gamma_i}{d\omega dE_{\nu} } = R \sum_{j} 
\int_{0}^1 dx_2 H_{ij} (\frac{\omega }{\omega_m },
\frac{ E_{\nu} }{\omega_m }, x_2)
\left(|c_{ij}^2| - \delta_M  \frac{3}{8} \frac{m_i m_j} {\omega_m^2 x_1 x_2} \Re(c_{ij}^2) \right)
\,,
\end{eqnarray}
where $\delta_M = 1$ for the Majorana neutrino and
 $\delta_M = 0$ for the Dirac neutrino.

Detection probability of a single neutrino event is
 estimated by the factor $\sigma n_N L$
where $n_N$ is the nucleon number density and $L$
is the detector's size along the neutrino beam.
The cross section is of order $10^{-39} \sim 10^{-38}$cm$^2$
for a 1 GeV neutrino, which gives
$\sigma n_N L \sim 10^{-11} \sim 10^{-10}$ (or smaller
depending on lower detected neutrino energy) for 
a single neutrino detection  using $\sim$100 m detector size.
One should multiply this factor  to  rate numbers shown in the presented figures,
in order to derive actual detection rates
of double events.

\begin{figure*}[htbp]
 \begin{center}
 \epsfxsize=0.6\textwidth
 \centerline{\epsfbox{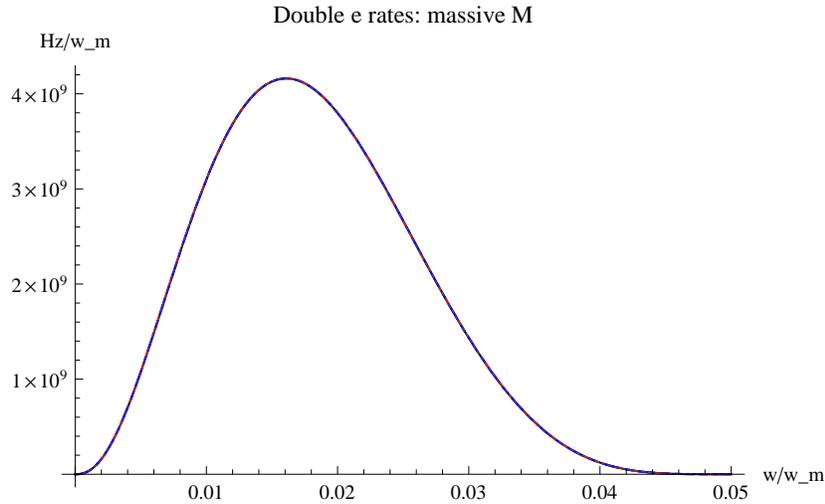}} \hspace*{\fill}
   \caption{Double spectral rates of  photon with a detected  $\nu_e$
of fixed energy $0.95 \omega_m = 9.5$MeV:  with the  smallest neutrino masses 5, 10,20, 50 meV's
all degenerate in the resolution of this figure.
All  cases for the Majorana NH of vanishing CPV paremeters,.
$N|\rho_{eg}(0)|^2 = 10^8,
\gamma_{pe}= 100 {\rm MHz},
\rho =4 {\rm km}$ and
$\gamma = 5 \times 10^3,
\epsilon_{eg} = 1{\rm keV},
\epsilon_{pe} = \epsilon_{eg}/10 = 0.1{\rm keV}$ are assumed.
}
\label{mdkev 3 massive rates}
 \end{center} 
\end{figure*}

To obtain  realistic absolute rates for measurements,
it is necessary to use a level spacing larger than of order keV,
as is evident from rate numbers in  Fig(\ref{renp kev rate contour}) and Fig(\ref{mdkev 3 massive rates}).
Since we consider $0(1)$keV energy range for the level spacing,
$\epsilon_{eg}$, the maximum neutrino energy $2\epsilon_{eg}\gamma$
is at most of order 100 MeV for $\gamma < O(10^4)$.
It is thus practical to consider double detection of combinations,
$\gamma + \nu_e$ and $\gamma + \bar{\nu}_e$,
since the charged current reactions $\nu_{\mu} \rightarrow \mu$ require neutrino energies much larger
than the muon mass.
Thus, the relevant spectrum rate is
\begin{eqnarray}
&&
\sum_i |U_{ei}|^2 \frac{d^2\Gamma_i}{d\omega dE_{\nu} } 
\,.
\end{eqnarray}
Neutrino detectors should be placed next to the photon
detector at the synchrotron site to make easier coincidence experiments.

\begin{figure*}[htbp]
 \begin{center}
 \epsfxsize=0.6\textwidth
 \centerline{\epsfbox{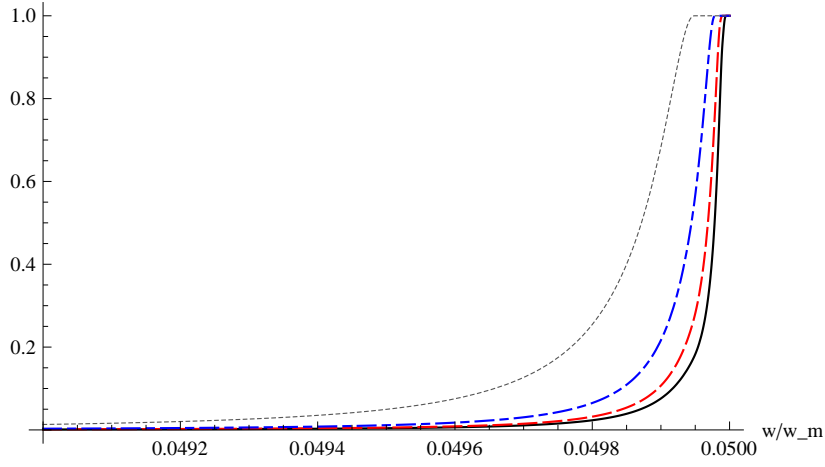}} \hspace*{\fill}
   \caption{Deviation of the rate ratio to
the three zero mass case from unity
for $\omega_m = 10$MeV:
5 meV in solid black, 10 meV in dashed red, 20 meV in dash-dotted blue,
and 50 meV in dotted black,
assuming the same set of parameters as in Fig(\ref{mdkev 3 massive rates}).
The integrated deviation is $4\times 10^{-5}$ for the 5 meV case,
while the double production rate in the threshold region of visible deviation is 
$\sim 2.5 \times 10^5$Hz.
}
\label {mdkev 3 massive difference}
 \end{center} 
\end{figure*}

The most promising case is to take
the neutrino energy near the maximally allowed value $\omega_m = 2 \epsilon_{eg} \gamma$,
as shown for $\omega_m=100$MeV 
in Fig(\ref{mdkev 3 massive rates}) and Fig(\ref{mdkev 3 massive difference})
in which the sensitivity to the smallest neutrino mass to of order
5 meV is indicated.
The smallest neutrino mass range can be explored 
by devising a  plot of deviation from the unity of
the ratio relative to the three massless neutrino
production rate,
$(d\Gamma(0) - d\Gamma(m_0)\,)/ d\Gamma(0)$,
which is shown for a few
choices of the smallest neutrino mass $m_0$ in Fig(\ref{mdkev 3 massive difference}).
The photon spectrum data may be obtained at different fixed neutrino energies, which
make analysis of absolute mass determination easier.
If calculated double detection rates are too small for a given detector design,
one should think of a machine construction of larger boost factor $\gamma$.
An increase of $\gamma$, for instance by 3, raises detection rates
by a large factor,  700 in this example.
There is no sensitivity to CP violating phases,
unless two neutrinos are simultaneously detected.

Let us compare the process of beam RENP discussed here with
RENP of SPAN project \cite{renp overview}.
A marked difference in experimental methods is
that the trigger laser irradiation is necessary for SPAN, while
it is unnecessary for the beam RENP.
Accordingly, the macro-coherence is required for SPAN, while
it is only the coherence at the single ion level in the beam RENP.
In terms of measurable quantities
SPAN has a better sensitivity for the smallest neutrino 
mass determination because of many thresholds present,
while the threshold rise is hardly visible here.
It would be instructive to write parameter dependence
of rates in two cases.
Disregarding common factors such as $G_F^2  N |\rho_{eg}(0)|^2$,
the rate in the quantum beam scales with 
$\sim 10^{-10} \gamma^6 \gamma_{pe} (\epsilon_{eg}/{\rm keV})^4$, while
SPAN RENP scales with $\sim 1 (n/10^{21}{\rm cm}^{-3})^2\gamma_{pg} ({\rm eV}/\epsilon_{eg})^2$
in the same unit.
We took the radius factor of $\sqrt{\rho \epsilon_{eg}} = O(10^5)$ for this comparison.

The important item for future R and D works
is fabrication of (1) intense lasers with high quality to obtain large
excited atom density $n$ for SPAN,
and (2) realizing heavy ion circulation with high coherence
$\rho_{eg}$ for its RENP.
In both cases a high coherence, either  at the macroscopic level
or at the single ion level, is of crucial importance for
further developments.

In summary,
radiative neutrino pair emission from circulating excited heavy ions
is  useful for determination of 
the smallest neutrino mass down to 5 meV level,
if a high coherence can be achieved in the quantum heavy
ion beam.
RENP from circulating heavy ions is complementary to
the neutrino pair beam, and two of them combined may
give a comparative perspective to SPAN project.

\vspace{0.5cm}
The author should like to thank N. Sasao for
valuable discussions on various experimental aspects
of the process discussed in the present work.
This research was partially supported by Grant-in-Aid for Scientific
Research on Innovative Areas "Extreme quantum world opened up by atoms"
(21104002) from the Ministry of Education, Culture, Sports, Science, 
and Technology, and JSPS KAKENHI Grant Number 15H02093.

\end{document}